\documentclass{optica-article}

\journal{opticajournal} 

\articletype{Research Article}

\usepackage{lineno}

\usepackage{hyperref}
\usepackage{float}
\usepackage{gensymb}
\usepackage{graphicx}
\usepackage{subcaption}
\usepackage{caption}
\usepackage{microtype}
\usepackage{booktabs}
\usepackage[inter-unit-product={\;},output-decimal-marker={.},per-mode=symbol-or-fraction,sticky-per,exponent-product = \cdot, output-product = \cdot,separate-uncertainty=true]{siunitx}
\usepackage[titletoc]{appendix}
\usepackage[wby]{callouts}
\usepackage{amsfonts}

\begin{document}

\title{A practical guide to loss measurements using the Fourier transform of the transmission spectrum}

\author{Hannah Thiel,\authormark{1,*} Bianca Nardi,\authormark{1} Alexander Schlager,\authormark{1} Stefan Frick,\authormark{1} and Gregor Weihs\authormark{1}}

\address{\authormark{1}Institut f\"{u}r Experimentalphysik, Universit\"{a}t Innsbruck, Technikerstra{\ss}e 25, 6020 Innsbruck, Austria}

\email{\authormark{*}hannah.thiel@uibk.ac.at} 

\begin{abstract*} 
Analyzing the internal loss characteristics and multimodedness of (integrated) optical devices can prove difficult.
One technique to recover this information is to Fourier transform the transmission spectrum of optical components.
This article gives instruction on how to perform the transmission measurement, prepare the data, and interpret the Fourier spectrum.
Our guide offers insights into the influence of sampling, windowing, zero padding as well as Fourier spectrum peak heights and shapes which are previously neglected in the literature but have considerable impact on the results of the method.
For illustration, we apply the method to a Bragg-reflection waveguide.
We find that the waveguide is multimodal with two modes having very similar group refractive indices but different optical losses.
\end{abstract*}

\section{Introduction}
In any optics or photonics system detailed knowledge of the optical loss of individual components is a prerequisite to optimizing performance.
This could be to stay competitive on the market, to preserve light coming from weak sources, as is often the case in biosensing~\cite{Singh2023,Cuesta2022} and whenever light cannot simply be amplified like light signals transmitted via free-space links~\cite{Bloom2003} or quantum states of light~\cite{Wootters1982}.
Quantifying these losses, however, can prove difficult; especially in integrated semiconductor devices.
The internal losses of any component tend to superimpose with the reflectivity of input and output facets, additional cavities forming between optical components, probe laser stability, and coupling and detector efficiencies.
A way of determining the true internal optical propagation loss of a component is a Fabry-Perot measurement.

It can be performed when the component undergoing testing acts as a cavity, such as a waveguide with cleaved facets.
When scanning the input wavelength, a fringe pattern arises at the output that originates from interference of the light inside the cavity.
The visibility of the Fabry-Perot fringes reveals the cavity loss, and therefore, propagation loss in a simple cavity~\cite{Feuchter1994}.
However, if the component in question is multimodal, structurally more complex or has fabrication imperfections, the fringe pattern becomes challenging to interpret.

An elegant way of analyzing the Fabry-Perot fringes in those cases is to perform a Fourier transform from the angular wavenumber $k$ to the optical path length $d$.
The Fourier spectrum contains information about the loss or gain in the medium, designed or unintentional defects and the modal landscape including group refractive indices.
The method was made popular by Hofstetter et al.~\cite{Hofstetter1998} as a way to measure losses in semiconductor lasers and has since been expanded to, among others, study defects in laser diodes~\cite{Lambkin2004}, the internal cavities introduced by tapers in photonic crystals~\cite{Talneau2003}, the single- or multi-modedness of waveguides~\cite{Pergande2010} and to modally resolve loss in waveguides~\cite{Pressl2015}.

While the theory is well developed and analysis seems straightforward, the actual measurement and its interpretation need to be done with care.
The resulting resonances and their magnitude depend sensitively on the implementation of the method.
We find that the necessary details are seldom described in publications.
This is why we would like to provide a practical guide for those wishing to apply the method in the lab.

To this end, we first introduce the basics of Fabry-Perot fringes, their Fourier transform, and the elements visible in the Fourier spectrum.
The practical guide explains how exactly to take the measurement and how to prepare the data for Fourier analysis paying attention to details like zero padding and windowing.
We detail the interpretation of the Fourier spectrum looking at peak shapes and heights and possible contributions from multiple modes.
Following this, we show how to obtain the propagation loss coefficients and group refractive indices from the peaks in the Fourier spectrum.
Finally, as a demonstration, we apply the method to an AlGaAs Bragg-reflection waveguide, a type of ridge waveguide used mainly as a source for photon pairs produced via parametric down-conversion~\cite{Pressl2018}. 

\section{The Fourier transform of a transmission spectrum}
A mathematical description of the transmission through a cavity and its Fourier transform has been given by Hofstetter and others and will not be repeated here~\cite{Hofstetter1997,Hofstetter1998,Lambkin2004}.
To perform the measurement it is sufficient to understand the following picture:

Consider light being reflected back and forth in a cavity.
In integrated photonics, this cavity could be made up of a waveguide with cleaved facets with moderate reflectivity.
The light will interfere destructively or constructively depending on its wavelength and the optical length of the cavity.
This interference manifests in the Fabry-Perot fringes measurable as light intensity coupled out of the cavity when the wavelength of the input light is scanned.

\begin{figure}[ht]
	\captionsetup{singlelinecheck = false, justification=raggedright}
	\centering
	\begin{annotate}{\includegraphics[width=3in]{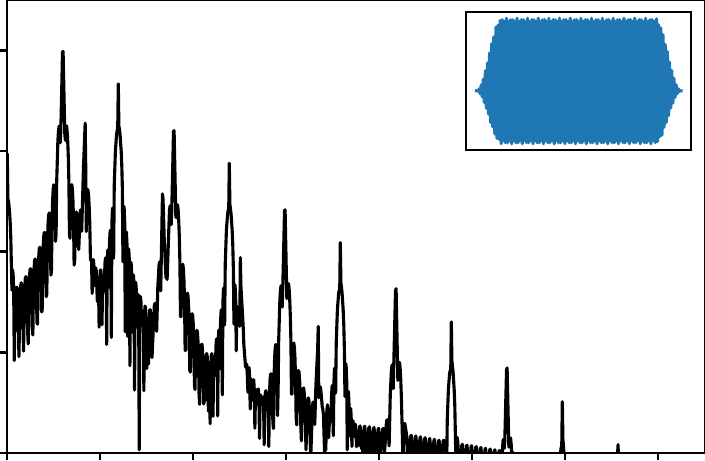}}{1}
		\draw [dashed][line width=0.5mm, orange ] (-3.5,2.25) -- (2.85,-2.3) node [midway] {};
		\draw [dashed][line width=0.5mm, red ] (-3.5,1.75) -- (1,-2.25) node [right] {};
		\draw[very thick,black] (-1.6,-2.8) node[anchor=north west]{Optical Length $d$ (mm)};
		\draw[very thick,black] (-3.93,-2.5) node[anchor=north west]{0};
		\draw[very thick,black] (-2.93,-2.5) node[anchor=north west]{2};
		\draw[very thick,black] (-1.93,-2.5) node[anchor=north west]{4};
		\draw[very thick,black] (-0.93,-2.5) node[anchor=north west]{6};
		\draw[very thick,black] (0.08,-2.5) node[anchor=north west]{8};
		\draw[very thick,black] (1,-2.5) node[anchor=north west]{10};
		\draw[very thick,black] (2,-2.5) node[anchor=north west]{12};
		\draw[very thick,black] (3,-2.5) node[anchor=north west]{14};
		\draw[very thick,black] (-1.3,3.4) node[anchor=north west]{Group Index};
		\draw[very thick,black] (-3.92,2.35) node[anchor=north west, rotate=90]{-};
		\draw[very thick,black] (-2.7,2.35) node[anchor=north west, rotate=90]{-};
		\draw[very thick,black] (-1.5,2.35) node[anchor=north west, rotate=90]{-};
		\draw[very thick,black] (-0.3,2.35) node[anchor=north west, rotate=90]{-};
		\draw[very thick,black] (0.9,2.35) node[anchor=north west, rotate=90]{-};
		\draw[very thick,black] (2.1,2.35) node[anchor=north west, rotate=90]{-};
		\draw[very thick,black] (3.3,2.35) node[anchor=north west, rotate=90]{-};
		\draw[very thick,black] (-3.92,3) node[anchor=north west]{0};
		\draw[very thick,black] (-2.7,3) node[anchor=north west]{5};
		\draw[very thick,black] (-1.6,3) node[anchor=north west]{10};
		\draw[very thick,black] (-0.4,3) node[anchor=north west]{15};
		\draw[very thick,black] (0.8,3) node[anchor=north west]{20};
		\draw[very thick,black] (2,3) node[anchor=north west]{25};
		\draw[very thick,black] (3.2,3) node[anchor=north west]{30};
		\draw[very thick,black] (-4.2,-2) node[anchor=south west, rotate=90]{Spectral Amplitude (a.u.)};
		\draw[very thick,black] (-4.4,-2.65) node[anchor=south west]{\(10^0\)};
		\draw[very thick,black] (-4.4,-1.55) node[anchor=south west]{\(10^2\)};
		\draw[very thick,black] (-4.4,-0.45) node[anchor=south west]{\(10^4\)};
		\draw[very thick,black] (-4.4,0.63) node[anchor=south west]{\(10^6\)};
		\draw[very thick,black] (-4.4,1.73) node[anchor=south west]{\(10^8\)};
		\draw[very thick,black] (0.9,0.4) node[anchor=south west]{1520nm};
		\draw[very thick,black] (2.55,0.4) node[anchor=south west]{1550nm};
		\draw[very thick,black] (1.15,0.65) node[anchor=north west, rotate=90]{-};
		\draw[very thick,black] (3.37,0.65) node[anchor=north west, rotate=90]{-};
		\draw[very thick,black] (1.2,0.8) node[anchor=south west, rotate=90]{Transm.};
	\end{annotate}
    \caption{Fourier spectrum of a simulated transmission spectrum (inset) for two modes with group indices $n_\mathrm{red}=3.5$ and $n_\mathrm{orange}=2.5$ and loss coefficients $\alpha_\mathrm{red}=1.30\,\mathrm{mm}^{-1}$ and $\alpha_\mathrm{orange}=0.08\,\mathrm{mm}^{-1}$. The peaks belonging to one mode are separated by the optical cavity length which corresponds to a physical cavity length of 1.5\,mm.}
    \label{fig:FourierSpec}
\end{figure}
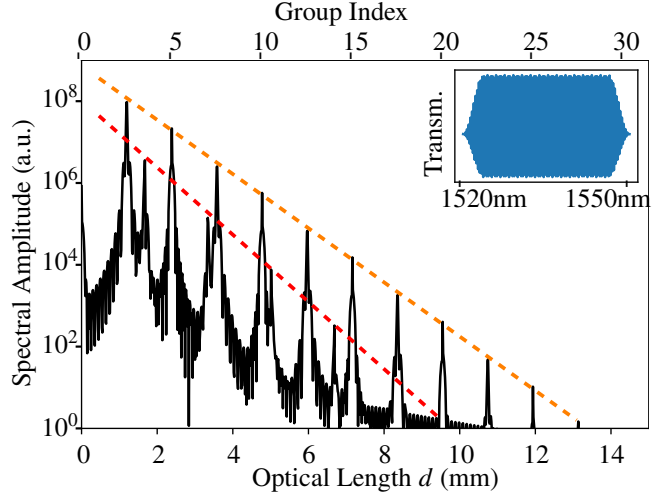

A propagating mode acquires phase shifts as it travels through the cavity.
When the Fabry-Perot fringes are Fourier transformed, these phase shifts manifest as signals at different values of the optical pathlength $d$.
For instance, every time light is reflected at the facets, it experiences a phase shift that sets apart light having travelled fewer or more passes through the cavity.
This is characterized by peaks in the Fourier spectrum spaced at integer multiples of the optical cavity length $L_\mathrm{opt}$, as showcased in Figure~\ref{fig:FourierSpec}.
The optical cavity length is proportional to the physical cavity length $L$ via a factor of $n/\pi$, where $n$ is the refractive index of the material, or, more specifically, the group index of the travelling mode~\footnote{For the definition of the optical cavity length we follow the notation introduced by Hofstetter et al. in References~\cite{Hofstetter1997} and~\cite{Hofstetter1998}.}.
The peaks decrease in amplitude as less of the light manages further passes.
This amplitude decrease depends on the reflectivity of the waveguide facets and the losses within the waveguide material.
Therefore, if the reflectivity $R$ of the facets is known and the ratio $\tilde{R}$ of subsequent peak heights is gathered from the Fourier spectrum, the intrinsic waveguide propagation loss coefficient
\begin{equation}
	\alpha = -\frac{1}{L}\ln\left(\frac{\tilde{R}}{R}\right)
	\label{eqn:Alpha}
\end{equation}
can be calculated via the formula derived in Ref.~\cite{Hofstetter1997}.
If the facet reflectivity is not known, it is also possible to determine it together with the loss coefficient.
In this case, the measurement must be done for multiple identical samples only differing in length~\cite{Pressl2015}.

In addition to the phase shifts upon reflection from the facets, each mode acquires a phase shift dependent on its group index during propagation.
Hence, instead of individual, equally spaced peaks, the Fourier spectrum features bunches of peaks each of which can be assigned to a mode with an individual optical cavity length.
The Fourier spectrum therefore contains information about the number of excited modes in the cavity, their relative strengths and their respective optical cavity lengths.
From this, one can calculate the travel times within the cavity and the group indices.

\section{Practical Guide: measurement, Fourier transform and analysis}
Knowing how powerful a tool the Fourier transform of a Fabry-Perot spectrum is, we can now move on to the practical guide.
Here, we take a look at the influence of sampling and data preparation.
We then Fourier transform the transmission spectrum and analyze the Fourier spectrum paying attention to peak positions, heights, and shapes.
Our protocol uses the fast Fourier transform contained in the NumPy library in Python specifically, but the results should hold for any established fast Fourier transform (FFT) algorithm.

\subsection{Influence of sampling}
The peak positions and amplitudes in the Fourier spectrum depend critically on how exactly the measurement is taken.
The first step is to record the transmission spectrum.

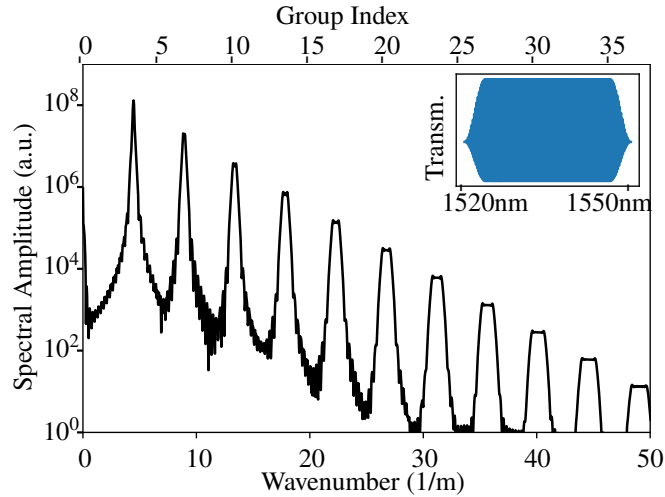
\begin{figure}[ht]
	\captionsetup{singlelinecheck = false, justification=raggedright}
	\centering
	\begin{annotate}{\includegraphics[width=3in]{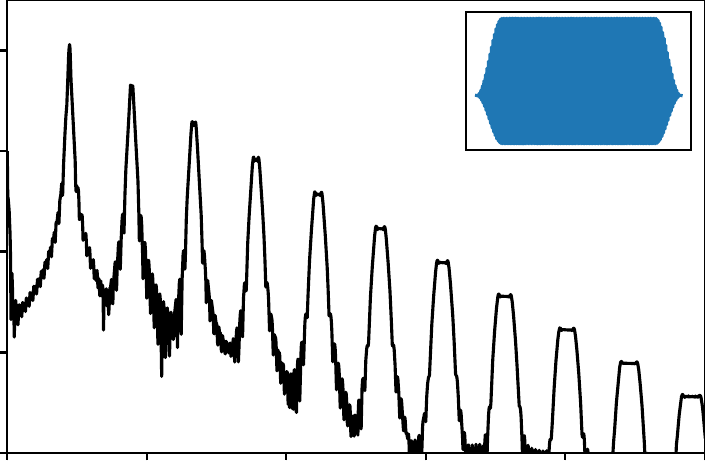}}{1}
		\draw[very thick,black] (-1.5,-2.8) node[anchor=north west]{Wavenumber (1/m)};
		\draw[very thick,black] (-3.93,-2.5) node[anchor=north west]{0};
		\draw[very thick,black] (-2.5,-2.5) node[anchor=north west]{10};
		\draw[very thick,black] (-1,-2.5) node[anchor=north west]{20};
		\draw[very thick,black] (0.5,-2.5) node[anchor=north west]{30};
		\draw[very thick,black] (2,-2.5) node[anchor=north west]{40};
		\draw[very thick,black] (3.5,-2.5) node[anchor=north west]{50};
		\draw[very thick,black] (-1.3,3.4) node[anchor=north west]{Group Index};
		\draw[very thick,black] (-3.92,2.35) node[anchor=north west, rotate=90]{-};
		\draw[very thick,black] (-2.9,2.35) node[anchor=north west, rotate=90]{-};
		\draw[very thick,black] (-1.9,2.35) node[anchor=north west, rotate=90]{-};
		\draw[very thick,black] (-0.9,2.35) node[anchor=north west, rotate=90]{-};
		\draw[very thick,black] (0.1,2.35) node[anchor=north west, rotate=90]{-};
		\draw[very thick,black] (1.1,2.35) node[anchor=north west, rotate=90]{-};
		\draw[very thick,black] (2.1,2.35) node[anchor=north west, rotate=90]{-};
		\draw[very thick,black] (3.1,2.35) node[anchor=north west, rotate=90]{-};
		\draw[very thick,black] (-3.92,3) node[anchor=north west]{0};
		\draw[very thick,black] (-2.9,3) node[anchor=north west]{5};
		\draw[very thick,black] (-2,3) node[anchor=north west]{10};
		\draw[very thick,black] (-1,3) node[anchor=north west]{15};
		\draw[very thick,black] (0,3) node[anchor=north west]{20};
		\draw[very thick,black] (1,3) node[anchor=north west]{25};
		\draw[very thick,black] (2,3) node[anchor=north west]{30};
		\draw[very thick,black] (3,3) node[anchor=north west]{35};
		\draw[very thick,black] (-4.2,-2) node[anchor=south west, rotate=90]{Spectral Amplitude (a.u.)};
		\draw[very thick,black] (-4.4,-2.65) node[anchor=south west]{\(10^0\)};
		\draw[very thick,black] (-4.4,-1.55) node[anchor=south west]{\(10^2\)};
		\draw[very thick,black] (-4.4,-0.45) node[anchor=south west]{\(10^4\)};
		\draw[very thick,black] (-4.4,0.63) node[anchor=south west]{\(10^6\)};
		\draw[very thick,black] (-4.4,1.73) node[anchor=south west]{\(10^8\)};
		\draw[very thick,black] (0.9,0.4) node[anchor=south west]{1520nm};
		\draw[very thick,black] (2.55,0.4) node[anchor=south west]{1550nm};
		\draw[very thick,black] (1.15,0.65) node[anchor=north west, rotate=90]{-};
		\draw[very thick,black] (3.37,0.65) node[anchor=north west, rotate=90]{-};
		\draw[very thick,black] (1.2,0.8) node[anchor=south west, rotate=90]{Transm.};
	\end{annotate}
    \caption{Fourier spectrum of simulated transmission spectrum (inset). Measuring the transmission at equally spaced wavelength steps leads to broadened peaks, while equally spaced $k$-values result in well defined peaks, as seen in Figure~\ref{fig:Windowing} (bottom). }
    \label{fig:Plateaus}
\end{figure}

\textbf{Measurement:}
The FFT algorithm requires equally spaced input values, hence we record the transmitted power in equally spaced steps.
It is usually recorded as a function of wavelength as this is the turning knob available for most lasers.
However, after Fourier transforming, this leads to broadened peaks because a signal that is periodic in  angular wavenumber $k$ is sampled in equally-spaced wavelength steps.
This means that the frequency is estimated to be lower (higher) where the wavelength steps correspond to larger (smaller) steps in $k$-space.
The resulting peaks in the Fourier spectrum are smeared across a range of x-values, as can be seen in Figure~\ref{fig:Plateaus}.
One solution would be to employ the non-uniform discrete Fourier transform, for which a Python package called PyNUFFT exists~\cite{Lin2018}.
This form of FFT, however, is not as well established.
Therefore, for reliable results, we set the laser to equally spaced $k$-values (or their corresponding wavelengths) during the measurement, where the Fourier domain, again, is the optical path length $d$.

\textbf{Resolution:}
In addition to being equally spaced, the recorded data points should also afford a certain range and thus resolution in the Fourier spectrum.
On the one hand, if the data points are not dense enough, high spatial frequencies cannot be measured.
On the other hand, in order to keep the measurement time reasonably small, one needs to find a balance between density and range.
The density of data points should be increased as far as the measurement equipment allows and then the range increased until individual peaks are resolved.
However, one should be careful not to extend the measurement range to angular wavenumbers near the bandgap or other strong nonlinearities.
The existence of nonlinearities could also be a reason to limit the input laser power for the transmission measurement.
There is a tradeoff between precision and accuracy in the $d$-domain.
While a larger measurement range improves the precision of the Fourier transform, it also means that losses across a larger range of wavenumbers contribute to the loss coefficient.
This needs to be evaluated carefully for each case.

\subsection{Influence of data preparation}
\begin{figure}[ht]
	\captionsetup{singlelinecheck = false, justification=raggedright}
	\centering
	\begin{annotate}{\includegraphics[width=3in]{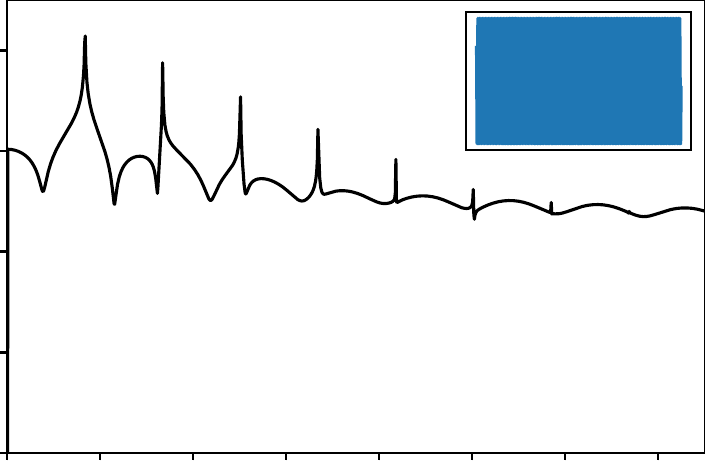}}{1}
		\draw[very thick,black] (-1.6,-2.8) node[anchor=north west]{Optical Length $d$ (mm)};
		\draw[very thick,black] (-3.93,-2.5) node[anchor=north west]{0};
		\draw[very thick,black] (-2.93,-2.5) node[anchor=north west]{2};
		\draw[very thick,black] (-1.93,-2.5) node[anchor=north west]{4};
		\draw[very thick,black] (-0.93,-2.5) node[anchor=north west]{6};
		\draw[very thick,black] (0.08,-2.5) node[anchor=north west]{8};
		\draw[very thick,black] (1,-2.5) node[anchor=north west]{10};
		\draw[very thick,black] (2,-2.5) node[anchor=north west]{12};
		\draw[very thick,black] (3,-2.5) node[anchor=north west]{14};
		\draw[very thick,black] (-1.3,3.4) node[anchor=north west]{Group Index};
		\draw[very thick,black] (-3.92,2.35) node[anchor=north west, rotate=90]{-};
		\draw[very thick,black] (-2.7,2.35) node[anchor=north west, rotate=90]{-};
		\draw[very thick,black] (-1.5,2.35) node[anchor=north west, rotate=90]{-};
		\draw[very thick,black] (-0.3,2.35) node[anchor=north west, rotate=90]{-};
		\draw[very thick,black] (0.9,2.35) node[anchor=north west, rotate=90]{-};
		\draw[very thick,black] (2.1,2.35) node[anchor=north west, rotate=90]{-};
		\draw[very thick,black] (3.3,2.35) node[anchor=north west, rotate=90]{-};
		\draw[very thick,black] (-3.92,3) node[anchor=north west]{0};
		\draw[very thick,black] (-2.7,3) node[anchor=north west]{5};
		\draw[very thick,black] (-1.6,3) node[anchor=north west]{10};
		\draw[very thick,black] (-0.4,3) node[anchor=north west]{15};
		\draw[very thick,black] (0.8,3) node[anchor=north west]{20};
		\draw[very thick,black] (2,3) node[anchor=north west]{25};
		\draw[very thick,black] (3.2,3) node[anchor=north west]{30};
		\draw[very thick,black] (-4.2,-2) node[anchor=south west, rotate=90]{Spectral Amplitude (a.u.)};
		\draw[very thick,black] (-4.4,-2.65) node[anchor=south west]{\(10^0\)};
		\draw[very thick,black] (-4.4,-1.55) node[anchor=south west]{\(10^2\)};
		\draw[very thick,black] (-4.4,-0.45) node[anchor=south west]{\(10^4\)};
		\draw[very thick,black] (-4.4,0.63) node[anchor=south west]{\(10^6\)};
		\draw[very thick,black] (-4.4,1.73) node[anchor=south west]{\(10^8\)};
		\draw[very thick,black] (0.9,0.4) node[anchor=south west]{1520nm};
		\draw[very thick,black] (2.55,0.4) node[anchor=south west]{1550nm};
		\draw[very thick,black] (1.15,0.65) node[anchor=north west, rotate=90]{-};
		\draw[very thick,black] (3.37,0.65) node[anchor=north west, rotate=90]{-};
		\draw[very thick,black] (1.2,0.8) node[anchor=south west, rotate=90]{Transm.};
	\end{annotate}
	\begin{annotate}{\includegraphics[width=3in]{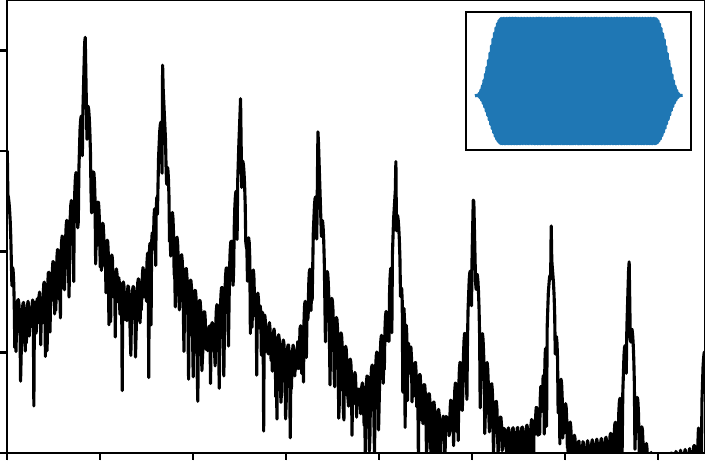}}{1}
		\draw[very thick,black] (-1.6,-2.8) node[anchor=north west]{Optical Length $d$ (mm)};
		\draw[very thick,black] (-3.93,-2.5) node[anchor=north west]{0};
		\draw[very thick,black] (-2.93,-2.5) node[anchor=north west]{2};
		\draw[very thick,black] (-1.93,-2.5) node[anchor=north west]{4};
		\draw[very thick,black] (-0.93,-2.5) node[anchor=north west]{6};
		\draw[very thick,black] (0.08,-2.5) node[anchor=north west]{8};
		\draw[very thick,black] (1,-2.5) node[anchor=north west]{10};
		\draw[very thick,black] (2,-2.5) node[anchor=north west]{12};
		\draw[very thick,black] (3,-2.5) node[anchor=north west]{14};
		\draw[very thick,black] (-1.3,3.4) node[anchor=north west]{Group Index};
		\draw[very thick,black] (-3.92,2.35) node[anchor=north west, rotate=90]{-};
		\draw[very thick,black] (-2.7,2.35) node[anchor=north west, rotate=90]{-};
		\draw[very thick,black] (-1.5,2.35) node[anchor=north west, rotate=90]{-};
		\draw[very thick,black] (-0.3,2.35) node[anchor=north west, rotate=90]{-};
		\draw[very thick,black] (0.9,2.35) node[anchor=north west, rotate=90]{-};
		\draw[very thick,black] (2.1,2.35) node[anchor=north west, rotate=90]{-};
		\draw[very thick,black] (3.3,2.35) node[anchor=north west, rotate=90]{-};
		\draw[very thick,black] (-3.92,3) node[anchor=north west]{0};
		\draw[very thick,black] (-2.7,3) node[anchor=north west]{5};
		\draw[very thick,black] (-1.6,3) node[anchor=north west]{10};
		\draw[very thick,black] (-0.4,3) node[anchor=north west]{15};
		\draw[very thick,black] (0.8,3) node[anchor=north west]{20};
		\draw[very thick,black] (2,3) node[anchor=north west]{25};
		\draw[very thick,black] (3.2,3) node[anchor=north west]{30};
		\draw[very thick,black] (-4.2,-2) node[anchor=south west, rotate=90]{Spectral Amplitude (a.u.)};
		\draw[very thick,black] (-4.4,-2.65) node[anchor=south west]{\(10^0\)};
		\draw[very thick,black] (-4.4,-1.55) node[anchor=south west]{\(10^2\)};
		\draw[very thick,black] (-4.4,-0.45) node[anchor=south west]{\(10^4\)};
		\draw[very thick,black] (-4.4,0.63) node[anchor=south west]{\(10^6\)};
		\draw[very thick,black] (-4.4,1.73) node[anchor=south west]{\(10^8\)};
		\draw[very thick,black] (0.9,0.4) node[anchor=south west]{1520nm};
		\draw[very thick,black] (2.55,0.4) node[anchor=south west]{1550nm};
		\draw[very thick,black] (1.15,0.65) node[anchor=north west, rotate=90]{-};
		\draw[very thick,black] (3.37,0.65) node[anchor=north west, rotate=90]{-};
		\draw[very thick,black] (1.2,0.8) node[anchor=south west, rotate=90]{Transm.};
	\end{annotate}
    \caption{Fourier spectra of a raw dataset (top) and of the same dataset multiplied with a Tukey window (bottom). Using a window removes the sinc that comes from Fourier transforming rectangular data. In both cases the data was augmented using a 200 entries long zero padding.}
    \label{fig:Windowing}
\end{figure}
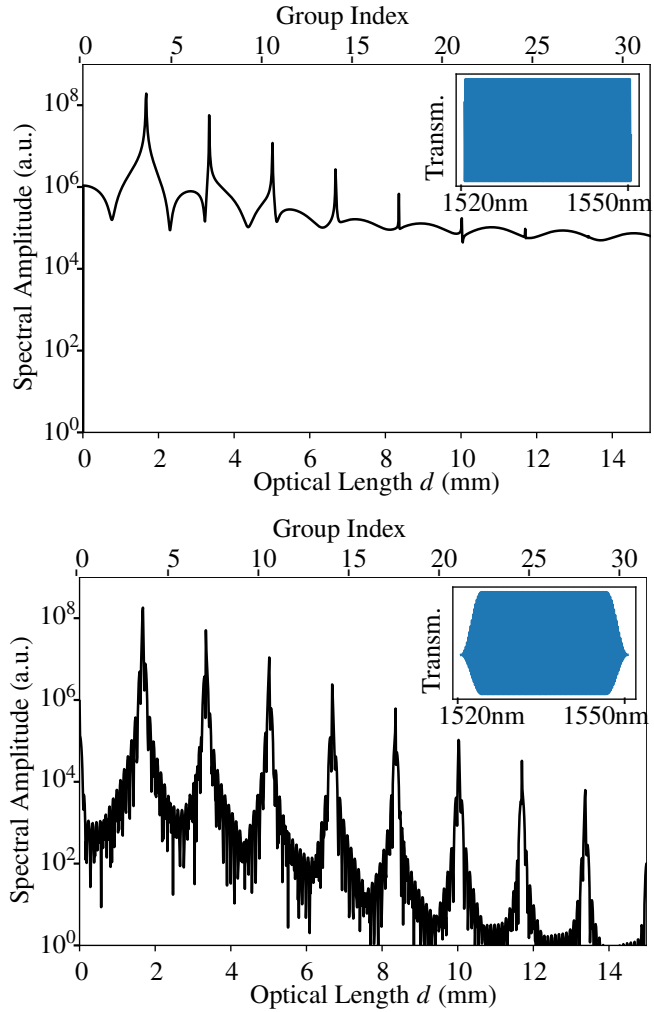

\textbf{Windowing:}
Once the data is recorded, it is advisable to multiply it with a window function.
The transmission data from complicated, multi-modal devices, especially, is unlikely to  match periodic boundary conditions.
The FFT algorithm interprets the start and endpoint of the dataset as if they were neighboring data points, which results in discontinuities for non-periodic datasets.
Not using any windowing is equivalent to applying a rectangular window, the Fourier transform of which approximates a sinc function.
This sinc superimposes with the signal in the Fourier spectrum distorting peak heights and therefore the magnitudes of modes.
To conserve the ``true" Fourier spectral amplitudes, we suggest using a window function with a wide main lobe.
Options are the Tukey window with a small shape parameter or the Flattop window~\cite{Harris1978,Heinzel2002}.
Figure~\ref{fig:Windowing} shows the comparison of a simulated dataset not treated with any window and that multiplied with a Tukey window (with a shape parameter of $\sim0.25$) before the Fourier transform.
One can see how the underlying sinc modifies the peak heights depending on where the side lobes sit relative to the peaks.
If, however, spectral resolution is important, for instance in order to discern two different modes, it is best to choose a window with a narrow main lobe, such as the popular Hann window~\cite{Heinzel2002}.

\textbf{Zero Padding:}
One of the parameters of the Python discrete FFT routine numpy.fft.rfft  is the number of input points the function will use.
Here, one can set a number larger than the input data set to add zero padding.
It is important to consider this when using the Fourier transformed Fabry-Perot spectrum as it influences peak heights and positions.
Zero padding is usually added to improve the numerical efficiency of the FFT algorithm or $d$-domain resolution.
A high number of added zeros may be useful for some applications where the $d$-axis position of the peaks needs to be known as precisely as possible, for instance in order to deduce group indices.
This aspect is not as important when trying to determine loss coefficients for which only amplitude accuracy is relevant, as shown in Equation~\ref{eqn:Alpha}.
Instead of their amplitude, one could also use the peak areas for the loss calculation.
However, this becomes impossible if the multimodedness of the structure causes overlapping peaks.

\begin{figure}[ht]
	\captionsetup{singlelinecheck = false, justification=raggedright}
	\centering
	\begin{annotate}{\includegraphics[width=3in]{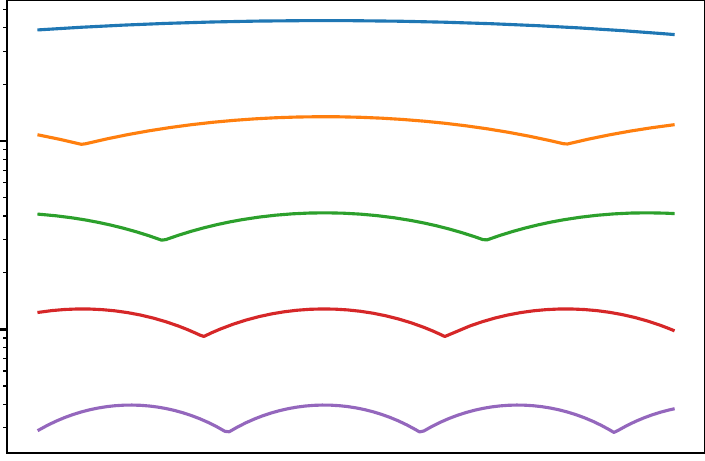}}{1}
		\draw [dashed][line width=0.3mm, black ] (-0.27,2.35) -- (-0.27,-2.25) node [right] {};
		\draw[very thick,black] (-1.5,-2.9) node[anchor=north west]{Number of Zero Pads};
		\draw[very thick,black] (-3.62,-2.5) node[anchor=north west]{0};
		\draw[very thick,black] (-1.4,-2.5) node[anchor=north west]{50};
		\draw[very thick,black] (0.8,-2.5) node[anchor=north west]{100};
		\draw[very thick,black] (3.1,-2.5) node[anchor=north west]{150};
		\draw[very thick,black] (-3.6,-2.65) node[anchor=north west, rotate=90]{-};
		\draw[very thick,black] (-1.3,-2.65) node[anchor=north west, rotate=90]{-};
		\draw[very thick,black] (1,-2.65) node[anchor=north west, rotate=90]{-};
		\draw[very thick,black] (3.3,-2.65) node[anchor=north west, rotate=90]{-};
		\draw[very thick,black] (-4.2,-2) node[anchor=south west, rotate=90]{Spectral Amplitude (a.u.)};
		\draw[very thick,black] (-4.4,-1.3) node[anchor=south west]{\(10^6\)};
		\draw[very thick,black] (-4.4,0.75) node[anchor=south west]{\(10^7\)};
  	\end{annotate}
    \caption{The spectral amplitudes of the first five peaks (different colors) are plotted as a function of zero padding length used in the Fourier transform. The peak height maxima coincide with those of the first peak because the simulation of the transmission spectrum was done for a single mode (black dashed line as a guide to the eye). These maxima are what the peak heights converge to for large zero paddings.}
    \label{fig:zeropadding}
\end{figure}
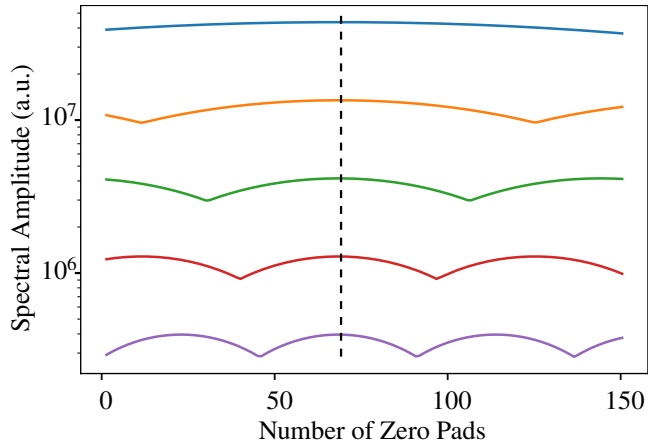

When the FFT is performed, the algorithm divides the $d$-axis into bins over which the signal is distributed.
The size and position of a bin depends on the number of data points used by the algorithm.
A maximum peak height is a result of the bin's spatial frequency coinciding with the exact frequency solution of the Fourier transform.
When a peak is separated into multiple bins, its spectral amplitude decreases accordingly.
This effect is visible even for a very low number of zeros added to a large data set with tens of thousands of data points, as illustrated in Figure~\ref{fig:zeropadding}.
We add zeros to the data set and, for each zero padding length, perform the FFT to find the peak heights.
One can see that the peak heights oscillate as a function of the padding length.
That means the Fourier spectra for these different zero paddings would yield vastly different loss coefficients.
Hence, choosing the appropriate zero padding length for each individual peak is crucial for the calculation of the loss coefficient.

The most rigorous way to determine the ideal zero padding would be to add infinitely many zeros to the measured data to resolve all frequencies and obtain the true height of each peak.
As this is numerically impossible, the next best thing is to add enough zeros for the peak heights to converge.
The values they converge to are their respective maxima.
However, the point at which they converge depends on the number and density of the data points measured and would have to be determined anew for each data set.
Therefore, to simplify the procedure, we scan through up to 150 added zeros and choose the smallest zero padding that maximizes the peak height for each of the peaks involved.
The heights of the different peaks do not necessarily reach a maximum at the same zero padding, as shown in Figure~\ref{fig:ZeropaddingBRW}.
We then extract each peak height from its individual spectrum to perform further analyses and calculations.

\subsection{Analysis}
\textbf{Peak Shapes and Modes:} When analyzing a Fourier spectrum, there are a couple of things to look out for.
The peak shape can vary from a clear single peak coming from a single mode to a single peak made up of contributions from multiple modes with similar group index or to a bundle of peaks corresponding to modes with different group indices.
If the peak is made up of a single mode, the ratios between subsequent peak heights will remain constant assuming a constant loss coefficient across the measured range.
It is then straightforward to use this ratio for the calculation of the loss coefficient.
In this case, one can also consider using the peak area instead of amplitude, as done in Ref.~\cite{Lambkin2004}.
When two modes with very similar group index propagate in the cavity, they might not be resolved and show up as signal in the same peak.
However, they might have a different loss coefficient which would then manifest as a change in peak height ratio between subsequent peaks.
This multimodedness can also be seen when plotting peak heights as a function of zero padding length; the height maxima of the first peak do not coincide with those of the later peaks.
To see this, compare Figures~\ref{fig:zeropadding} and~\ref{fig:ZeropaddingBRW}.
When there are multiple modes with distinct group indices, they will show up as separate peaks in the Fourier spectrum.
Each peak will have higher harmonics at multiples of their individual optical resonator length.
While the peak bundle might change appearance, the constituent modes are easily identifiable, as is the case in Figure~\ref{fig:FourierSpec}.

\textbf{Peak Height Uncertainty:} Once the contributions of the modes in the spectrum have been identified, the peak heights (or areas if applicable) can be determined from the Fourier spectrum.
Depending on what the final goal of the analysis is, it may be important to consider the uncertainties of the peak heights.
In Ref.~\cite{Eichstaedt2016}, Eichstaedt et al. present an open-source Python software tool which treats the propagation of uncertainties in the (inverse) discrete Fourier transform, also taking into account windowing and zero padding.
Another way of arriving at an estimate of the peak height uncertainty is to calculate the  peak heights in the Fourier spectra created with one more or one fewer zero pad than the ideal one.
We then calculate the difference to the peak heights obtained with the ideal zero padding.
Whichever difference is smaller can then be used as the uncertainty.

\textbf{Peak Height Ratios:} Finally, in order to determine the loss coefficient $\alpha$, we need to calculate the ratio of subsequent peak heights.
For a sample that acts as a single mode cavity with constant loss across the measurement range, the peak height ratios are constant.
Depending on the quality factor of the cavity, one can identify just a handful or many peaks in the Fourier spectrum.
As we move along the $d$-axis towards larger optical lengths, the peaks decrease in height and correspond to lower signal intensity.
They should therefore be weighted accordingly when determining the overall $\tilde{R}$.
If the sample is more complex, it might make sense to choose an individual weighting or even to only use two peaks and their peak height ratio for the calculations.
One such case is explained in detail in the following section.

\section{Example: Analysis of a Bragg-reflection waveguide}
To demonstrate the technique of Fourier transforming a transmission spectrum, we apply it to a Bragg-reflection waveguide (BRW).
Our BRWs are semiconductor devices made of AlGaAs that are used for the production of entangled photon pairs in the telecom C-band via parametric down-conversion from the NIR.
In order to be able to guide and phase-match the modes involved in the down-conversion process, the waveguide is made up of layers with different aluminum concentration.
A more detailed description of BRWs can be found in Refs.~\cite{Yeh1976,Helmy,Pressl2018}.
The complicated layer stack and large cross-section mean that the waveguide is multimodal.
Additionally, fabrication is challenging and imperfections reflect strongly in the loss coefficient~\cite{Thiel2023}.

\begin{figure}[ht]
	\captionsetup{singlelinecheck = false, justification=raggedright}
	\centering
	\begin{annotate}{\includegraphics[width=3in]{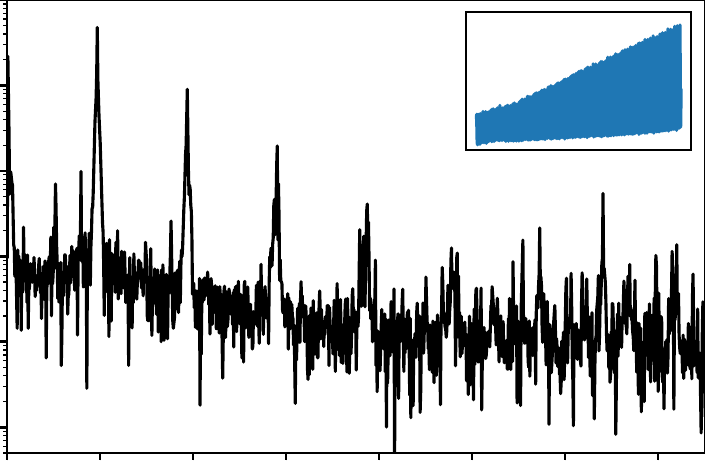}}{1}
		\draw[very thick,black] (-1.6,-2.8) node[anchor=north west]{Optical Length $d$ (mm)};
		\draw[very thick,black] (-3.93,-2.5) node[anchor=north west]{0};
		\draw[very thick,black] (-2.93,-2.5) node[anchor=north west]{2};
		\draw[very thick,black] (-1.93,-2.5) node[anchor=north west]{4};
		\draw[very thick,black] (-0.93,-2.5) node[anchor=north west]{6};
		\draw[very thick,black] (0.08,-2.5) node[anchor=north west]{8};
		\draw[very thick,black] (1,-2.5) node[anchor=north west]{10};
		\draw[very thick,black] (2,-2.5) node[anchor=north west]{12};
		\draw[very thick,black] (3,-2.5) node[anchor=north west]{14};
		\draw[very thick,black] (-1.3,3.4) node[anchor=north west]{Group Index};
		\draw[very thick,black] (-3.92,2.35) node[anchor=north west, rotate=90]{-};
		\draw[very thick,black] (-2.7,2.35) node[anchor=north west, rotate=90]{-};
		\draw[very thick,black] (-1.5,2.35) node[anchor=north west, rotate=90]{-};
		\draw[very thick,black] (-0.3,2.35) node[anchor=north west, rotate=90]{-};
		\draw[very thick,black] (0.9,2.35) node[anchor=north west, rotate=90]{-};
		\draw[very thick,black] (2.1,2.35) node[anchor=north west, rotate=90]{-};
		\draw[very thick,black] (3.3,2.35) node[anchor=north west, rotate=90]{-};
		\draw[very thick,black] (-3.92,3) node[anchor=north west]{0};
		\draw[very thick,black] (-2.7,3) node[anchor=north west]{5};
		\draw[very thick,black] (-1.6,3) node[anchor=north west]{10};
		\draw[very thick,black] (-0.4,3) node[anchor=north west]{15};
		\draw[very thick,black] (0.8,3) node[anchor=north west]{20};
		\draw[very thick,black] (2,3) node[anchor=north west]{25};
		\draw[very thick,black] (3.2,3) node[anchor=north west]{30};
		\draw[very thick,black] (-4.2,-2) node[anchor=south west, rotate=90]{Spectral Amplitude (a.u.)};
		\draw[orange] (-2.78,2.2) circle (3pt);
		\draw[orange] (-1.78,1.53) circle (3pt);
		\draw[orange] (-0.82,0.9) circle (3pt);
		\draw[orange] (0.15,0.3) circle (3pt);
		\draw[orange] (1.13,-0.23) circle (3pt);
		\draw[very thick,black] (-4.4,-2.36) node[anchor=south west]{\(10^0\)};
		\draw[very thick,black] (-4.4,-1.43) node[anchor=south west]{\(10^1\)};
		\draw[very thick,black] (-4.4,-0.54) node[anchor=south west]{\(10^2\)};
		\draw[very thick,black] (-4.4,0.4) node[anchor=south west]{\(10^3\)};
		\draw[very thick,black] (-4.4,1.3) node[anchor=south west]{\(10^4\)};
		\draw[very thick,black] (-4.4,2.25) node[anchor=south west]{\(10^5\)};
		\draw[very thick,black] (0.9,0.4) node[anchor=south west]{1520nm};
		\draw[very thick,black] (2.55,0.4) node[anchor=south west]{1550nm};
		\draw[very thick,black] (1.15,0.65) node[anchor=north west, rotate=90]{-};
		\draw[very thick,black] (3.37,0.65) node[anchor=north west, rotate=90]{-};
		\draw[very thick,black] (0.4,0.95) node[anchor=south west]{10\(\%\)};
		\draw[very thick,black] (0.4,1.9) node[anchor=south west]{20\(\%\)};
		\draw[very thick,black] (1,1.45) node[anchor=north west]{-};
		\draw[very thick,black] (1,2.4) node[anchor=north west]{-};	
  	\end{annotate}
    \caption{The Fourier transform of the telecom transmission spectrum through a BRW features individual peaks spaced at the optical cavity length. Adapted from~\cite{Thiel2023}.}
    \label{fig:FourierBRW}
\end{figure}
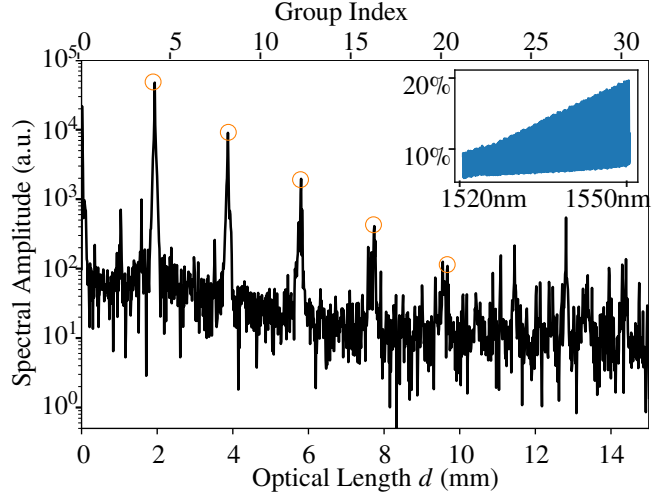

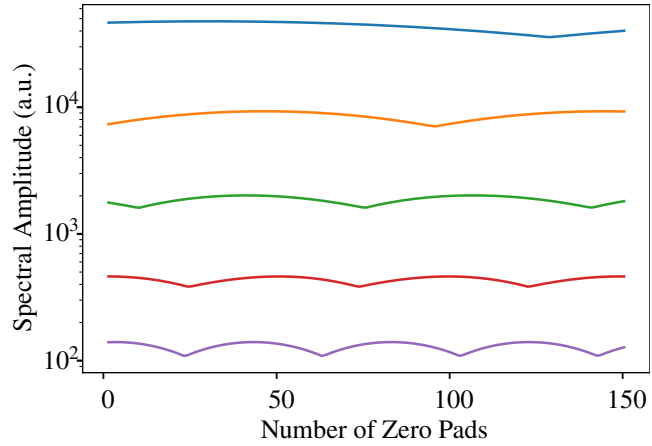
\begin{figure}[ht]
	\captionsetup{singlelinecheck = false, justification=raggedright}
	\centering
	\begin{annotate}{\includegraphics[width=3in]{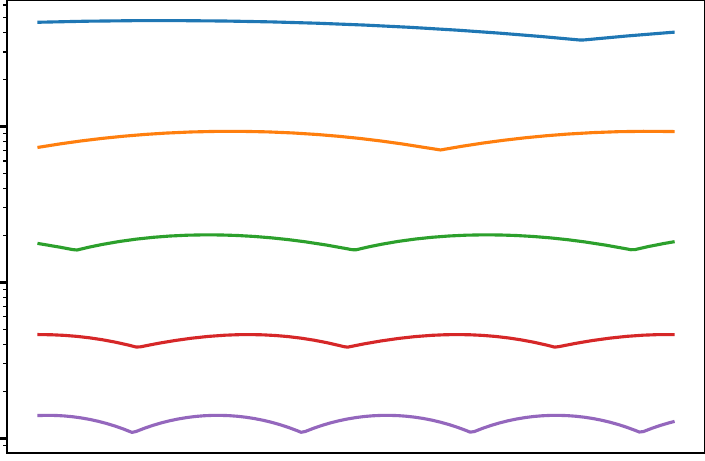}}{1}
		\draw[very thick,black] (-1.5,-2.9) node[anchor=north west]{Number of Zero Pads};
		\draw[very thick,black] (-3.62,-2.5) node[anchor=north west]{0};
		\draw[very thick,black] (-1.4,-2.5) node[anchor=north west]{50};
		\draw[very thick,black] (0.8,-2.5) node[anchor=north west]{100};
		\draw[very thick,black] (3.1,-2.5) node[anchor=north west]{150};
		\draw[very thick,black] (-3.6,-2.65) node[anchor=north west, rotate=90]{-};
		\draw[very thick,black] (-1.3,-2.65) node[anchor=north west, rotate=90]{-};
		\draw[very thick,black] (1,-2.65) node[anchor=north west, rotate=90]{-};
		\draw[very thick,black] (3.3,-2.65) node[anchor=north west, rotate=90]{-};
		\draw[very thick,black] (-4.2,-2) node[anchor=south west, rotate=90]{Spectral Amplitude (a.u.)};
		\draw[very thick,black] (-4.4,-2.5) node[anchor=south west]{\(10^2\)};
		\draw[very thick,black] (-4.4,-0.8) node[anchor=south west]{\(10^3\)};
		\draw[very thick,black] (-4.4,0.9) node[anchor=south west]{\(10^4\)};
  	\end{annotate}
    \caption{The spectral amplitudes of the first five peaks are plotted as a function of zero padding length used in the Fourier transform. The peak height maxima do not coincide for the  BRW. This points to multiple modes propagating in the cavity.}
    \label{fig:ZeropaddingBRW}
\end{figure}

We perform the Fourier analysis of the transmission spectrum for TE-polarized light in the telecom wavelength range.
To this end, we scan the wavelength of our tunable telecom laser (Santec TSL-710) in steps that correspond to wavenumber steps of around \(2.49\,\mathrm{m}^{-1}\) in the range from $1520-1550\, \mathrm{nm}$.
We apply a Tukey window with shape parameter 0.25 to the data and Fourier transform for different zero padding lengths.
Figure~\ref{fig:FourierBRW} shows the Fourier spectrum for one specific zero padding.
It features single peaks separated by the optical length of the cavity pointing to a single-mode operation or multi-mode operation where the modes have very similar group indices.
A plot of the peak heights as a function of zero padding length is shown in Figure~\ref{fig:ZeropaddingBRW}.

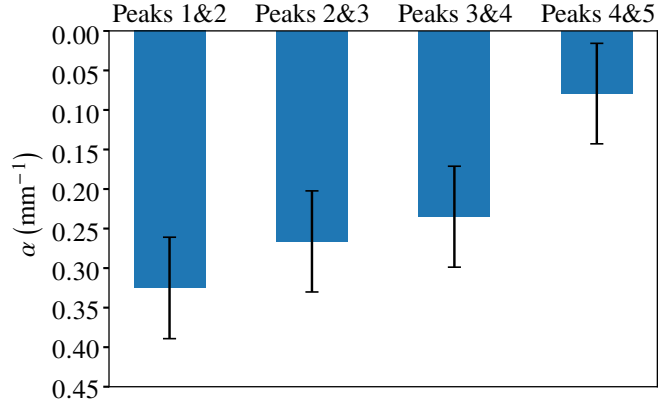
\begin{figure}[ht]
	\captionsetup{singlelinecheck = false, justification=raggedright}
	\centering
    \begin{annotate}{\includegraphics[width=2.9in]{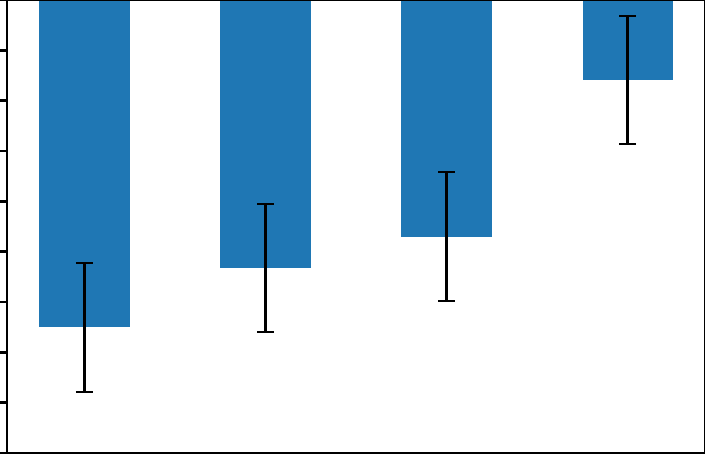}}{1}
    	\draw[very thick,black] (-3.7,2.9) node[anchor=north west]{Peaks 1\&2};
    	\draw[very thick,black] (-1.8,2.9) node[anchor=north west]{Peaks 2\&3};
    	\draw[very thick,black] (0.1,2.9) node[anchor=north west]{Peaks 3\&4};
    	\draw[very thick,black] (2,2.9) node[anchor=north west]{Peaks 4\&5};
    	\draw[very thick,black] (-4.35,-0.75) node[anchor=south west, rotate=90]{\(\alpha \left(\mathrm{mm}^{-1}\right)\)};
    	\draw[very thick,black] (-4.5,2.6) node[anchor=north west]{$0.00$};
    	\draw[very thick,black] (-4.5,2.078) node[anchor=north west]{$0.05$};
    	\draw[very thick,black] (-4.5,1.556) node[anchor=north west]{$0.10$};
    	\draw[very thick,black] (-4.5,1.034) node[anchor=north west]{$0.15$};
    	\draw[very thick,black] (-4.5,0.512) node[anchor=north west]{$0.20$};
    	\draw[very thick,black] (-4.5,-0.01) node[anchor=north west]{$0.25$};
    	\draw[very thick,black] (-4.5,-0.532) node[anchor=north west]{$0.30$};
    	\draw[very thick,black] (-4.5,-1.054) node[anchor=north west]{$0.35$};
    	\draw[very thick,black] (-4.5,-1.576) node[anchor=north west]{$0.40$};
    	\draw[very thick,black] (-4.5,-2.098) node[anchor=north west]{$0.45$};
    \end{annotate}   
    \caption{Loss coefficients $\alpha$ calculated from the ratios of neighboring peaks in the Fourier spectrum of a BRW. Adapted from~\cite{Thiel2023}.}
    \label{fig:AlphasBRW}
\end{figure}

We determine that a zero padding of $z_1=29$, $z_2=45$, $z_3=40$, $z_4=49$, and $z_5=2$ maximizes the peak height of the first, second, third, fourth, and fifth peak, respectively, and deduce from this the peak heights $h_1=4.7664(1)\cdot10^7$, $h_2=9.2927(3)\cdot10^6$, $h_3=2.0140(4)\cdot10^6$, $h_4=4.61784(6)\cdot10^5$, and $h_5=1.4013(6)\cdot10^5$.
Using the peak height ratios $\tilde{R}_{12}=h_2/h_1$, $\tilde{R}_{23}=h_3/h_2$, $\tilde{R}_{34}=h_4/h_3$, $\tilde{R}_{45}=h_5/h_4$, the physical cavity length of $L=1.80(5)\,\mathrm{mm}$, and the facet reflectivity of $R=0.35(4)$~\cite{Pressl2015}, we calculate the loss coefficients $\alpha$ using Equation~\ref{eqn:Alpha}.
The main contributions to the final uncertainty of the loss coefficient come from the systematic uncertainty of the cavity physical length and the facet reflectivity. The length of the BRW was measured using an optical microscope and calibration slide. The facet reflectivity was determined in an experiment involving multiple BRW samples with different length~\cite{Pressl2015}.
These uncertainties dominate over the statistical errors of the peak height ratios, which are smaller than $1.3\cdot10^{-4}$.
We calculate $\alpha$ separately for every neighboring peak pair to be able to analyze the multimodedness of the waveguide sample.
Looking at Figure~\ref{fig:AlphasBRW}, one can see that it decreases from \( \alpha_\mathrm{12}=0.33\left(6\right)\,\mathrm{mm}^{-1}\) between peaks one and two to \( \alpha_\mathrm{45}=0.08\left(6\right)\,\mathrm{mm}^{-1}\) between peaks four and five.
A non-constant loss coefficient means that at least two modes with very similar group indices propagate in the cavity.
The high loss coefficient of one mode dominates at the beginning, but its contribution diminishes for later peaks as it loses power.
We suspect that the field intensity of this mode is higher near the sidewall of the waveguide ridge.
It therefore experiences increased scattering from the sidewall imperfections.
Another mode, presumably closer to the core of the waveguide, benefits from lower scattering rates, survives more passes through the waveguides, and therefore features a lower loss coefficient.
The loss coefficient $\alpha_\mathrm{45}$ calculated from the latest peak ratio $\tilde{R}_{45}$ therefore approximates the loss coefficient of that less lossy mode.
We are thus able to analyze the loss characteristics of a multimodal waveguide structure, even if the group indices of the propagating modes are similar.

\section{Conclusion and Outlook}
We showed how to analyze the Fabry-Perot fringes in the transmission spectrum of multimodal or imperfect waveguide cavities using the Fourier transform.
The Fourier spectrum contains information about the loss characteristics and group refractive indices of propagating modes that can be extracted by following our practical guide.
We achieve satisfactory resolution by performing the transmission measurement at steps corresponding to densely and equally distributed wavenumbers $k$.
To prepare the data for the Fourier transform, we suggest a window with a wide main lobe like the Tukey window with small shape parameter.
For the peaks in the Fourier spectrum to reach their true heights, we find the ideal zero padding for each of the peaks individually.
Finally, after applying Python's NumPy FFT algorithm, we interpret the Fourier spectrum.
We find that the BRW analyzed is multimodal with two modes having very similar group refractive indices.
From the peak height ratios we can calculate the optical loss, which decreases from \( \alpha_\mathrm{12}=0.33\left(6\right)\,\mathrm{mm}^{-1}\) to \( \alpha_\mathrm{45}=0.08\left(6\right)\,\mathrm{mm}^{-1}\).
This means that the higher-loss mode dominates at few passes through the waveguide cavity while the lower-loss mode survives longer.
Detailed knowledge about the loss characteristics and modal landscape of the BRW helps us understand design flaws and fabrication imperfections.
The Fourier transform of the transmission spectrum can therefore be a useful tool for designers of any optics or photonics component.
Especially for integrated devices, this practical guide helps investigate internal losses separately from surrounding optical components.

\begin{backmatter}

\bmsection{Funding}
The authors acknowledge funding by the Uniqorn project (Horizon 2020 grant agreement no. 820474) and the BeyondC project (FWF project no. F7114).

\bmsection{Acknowledgments}
The authors thank Lukas Einkemmer for fruitful discussions and constructive criticism.

\bmsection{Author contributions}
Conceptualization, H.T., S.F., G.W.; Formal analysis, H.T., S.F.; Methodology,  H.T., S.F.; Investigation, H.T., A.S., B.N.; Software, H.T., A.S., S.F.; Supervision, S.F., G.W.; Writing - original draft, H.T.; Writing - review \& editing, All Authors; Funding acquisition, G.W.

\bmsection{Disclosures}
The authors declare no conflicts of interest.

\bmsection{Data Availability Statement}
The data that support the findings of this study are openly available at the following DOI: 10.5281/zenodo.7966624

\end{backmatter}

\bibliography{Optica-template}

\end{document}